\documentclass[]{article}
\usepackage{epsf}
\usepackage{graphicx,amssymb,lineno}

\title{Detecting the Most Unusual Part of Two and Three-dimensional Digital Images}

\author{
  Kostadin Koroutchev$^1$ and Elka Korutcheva$^{2,3}$\\
$^1${\small EPS, Universidad Aut\'onoma de Madrid, }\\
{\small Cantoblanco, Madrid, 28049, Spain}\\
{\small\it{k.koroutchev@uam.es}}\\
$^2${\small Depto. de F\'{\i}sica Fundamental, }\\
{\small Universidad Nacional de Educaci\'on a Distancia,}\\
{\small c/ Senda del Rey 9, 28080 Madrid, Spain}\\
$^3${\small G.Nadjakov Inst. Solid State Physics, }\\
{\small Bulgarian Academy of Sciences, Sofia, Bulgaria}
}

\date{}

\begin{document}
\maketitle
\begin{abstract}
The purpose of this paper is to introduce an algorithm that can detect the 
most unusual part of a digital image in probabilistic setting.
The most unusual part of a given shape is defined as a part of the 
image that has the maximal distance
to all non intersecting shapes with the same form.
The method is tested on two and three-dimensional images and
  has shown very good results without any predefined model. 
 A version of the method independent of 
the contrast of the image is considered and is found to be useful 
for finding the most unusual 
part (and the most similar part) of the image conditioned on 
given image.

The results can be
  used to scan large image databases, 
as for example medical databases.\\

\end{abstract}

\section{Introduction}

In this paper we are trying to find the most unusual part with 
predefined shape of a given image. If we consider an one-dimensional 
quasi-periodical image, as for example electrocardiogram (ECG), the most 
unusual parts with length about one second will be the parts that correspond
to 
rhythm abnormalities \cite{SAX}. Therefore they are of some interest. 
Considering two and three dimensional images, we can suppose that the most unusual 
part of the image can correspond to something interesting of the
image.

Recently we have presented an algorithm that can detect the most
  unusual part of a digital image, referring to two-dimensional images
  \cite{Buffalo}.
In fact the algorithm can be used in more that two dimensions.
 The present paper is an extension of this method in the case of three-dimensional images. 

Of course, if we have a clear mathematical model of what the interesting part
of the image can be, it would be probably better to build a mathematical model
that detects those unusual characteristics of the image part that are
interesting. However, as in the case of ECG, the part that we are looking for,
can not be defined by a clear mathematical model, or just the model can not be 
available. In such cases the most unusual part can be an interesting
 instrument for screening images.

To state the problem, we need first of all a definition of the term "most
unusual part". Let us chose some shape $S$ within the image $A$, 
that could contain that part
and let us denote the cut of the figure $A$ with shape $S$ and origin $\vec r$
by $ A_S(\vec \rho; \vec r) $, e.g. 
\[
A_S(\vec \rho; \vec r)\equiv S(\vec\rho) A(\vec\rho+\vec r), 
\]
where $\vec \rho$ is the in-shape coordinate vector, $\vec r$ 
is the origin of the 
cut $A_S$ and 
we used the characteristic function $S(.)$ of the shape $S$.
Further in this paper we will omit the arguments of $A_S$.
We can suppose that the most unusual part is the one
that has the largest distance with the rest of the cuts with the same
shape. 

Speaking mathematically, we can suppose that the most unusual part is 
located at the 
point $\vec r$, defined by:
\begin{equation}
\label{eq}
\vec r=\arg\max_{\vec r}\ \min_{\vec {r'}: |\vec {r'}-\vec r|>{\rm diam} (S)} 
|| A_S(\vec r)-A_S(\vec {r'})||.
\end{equation}
Here we assume that the shifts do not cross the border of the image. 
The norm $||.||$ is assumed to be $L_2$ norm\footnote{Similar results are achieved with $L_1$ norm. 
The algorithm was not tested with $L_{max}$ norm due to its extreme noise sensitivity. 
We use $L_2$ because of its relation with PSNR criteria that closely resembles the human subjective perception.}.

As the parts of an image that intersect significantly are similar, 
we do not allow the shapes located at  $r'$ and $r$ to intersect, avoiding 
this by the restriction on $r': |\vec {r'}-\vec r|>{\rm diam}(S)$.  

If we are looking for the part of the image to be unusual in a context of an
image 
database, we can assume that further restrictions on $r'$ can be added, for
example 
restricting the search to avoid intersection with several images.

The definition above can be interesting as a mathematical construction, but if we are looking for practical applications, it is too strict 
and does not correspond exactly to the intuitive notion of the interesting part as there can be several of them. Therefore the correct definition will be to find the outliers of the distribution of the distances $||.||$ between the blocks.

{  In $d$-dimensional space the figure with linear size $N$ 
has $N^d$ points and
if $||S||\ll ||A||$, in order to find 
deterministically the most unusual part, we need $N^d$
operations. }
This is unacceptable for large two dimensional images, 
and it is even worse in the case of 3D image databases. 
Therefore we are looking for an algorithm that provides an approximate 
solution of the problem and solves it within some probability limit
{  in acceptable execution time}.

As is defined above in Eq.(\ref{eq}), the problem is very similar to 
the problem of location of the nearest neighbor between the blocks. 
This problem has been studied in the literature, concerning Code Book 
and Fractal Compression \cite{fractal}. However, the problem of
 finding $\vec{r}$ in the above equation, without specifying $\vec {r'}$, 
as we show in the present paper, can be solved by using probabilistic methods 
avoiding slow calculations.


\section{The Method}

\subsection{Projections}

The problem of estimating the minima of Eq. (\ref{eq}) is complicated 
because the blocks are multidimensional. Therefore we can try to simplify 
the problem by projecting the block $B\equiv A_S(\vec r)$ in one 
dimension using some projection 
operator $X$. For this aim, we consider the following quantity:

\begin{equation}
\label{modul}
b=|X .B_1 - X . B| =|X.(B_1-B)|, \,\,\, |X|=1.
\end{equation}

The dot product in the above equation is the sum over all $\rho$-s:
\begin{equation}\label{projdef}
X.B\equiv\sum_{\vec \rho} X(\vec\rho) B(\vec\rho;\vec r). 
\end{equation}
If  $X$ is random, and uniformly distributed on the sphere of corresponding dimension, then the mean value 
of $b$ is proportional to $|B_1-B|$; $\langle b\rangle =c|B_1-B|$ and the coefficient $c$ depends only on the number of points of the block, that can be treated as its dimensionality, considering the projection operator. 
However, when the size of the block, e.g. its dimensionality increases, the two random vectors ($B_1-B$ and $X$) are close to orthogonal and the typical projection is small.
But if some block is far away from all the other blocks, then with some 
probability, the projection will be large. 
The method resembles that of Ref. \cite{indyk} for finding nearest neighbor.
\begin{figure}[t]
\begin{center}
\begin{minipage}{0.45\textwidth}
 \epsfysize=3.9cm
 \begin{center}
 \epsfbox{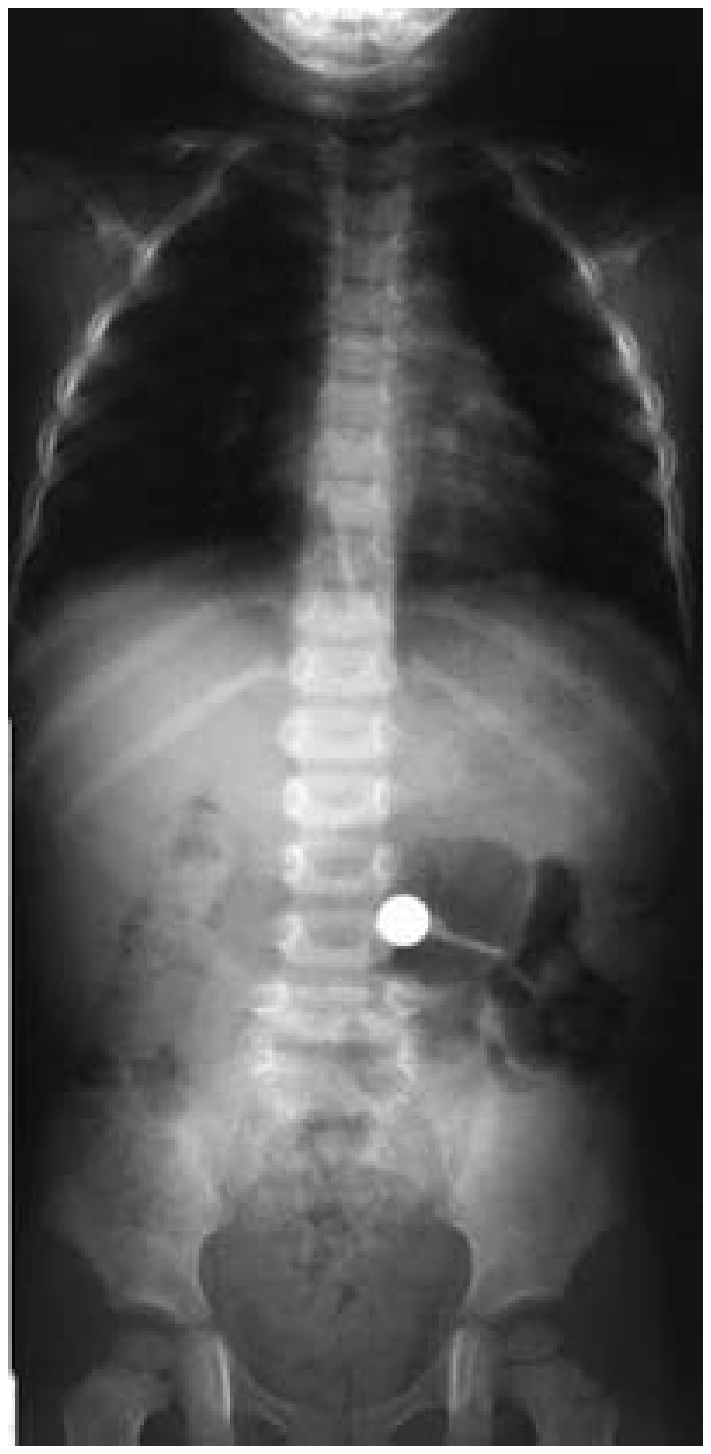}
 \end{center}
 \caption{The original test image. X-ray image of a person with ingested coin.
 }\label{figorig}
\end{minipage}
\hfill
\begin{minipage}{0.45\textwidth}
 \epsfysize 4.2cm 
 \begin{center}
 \epsfbox{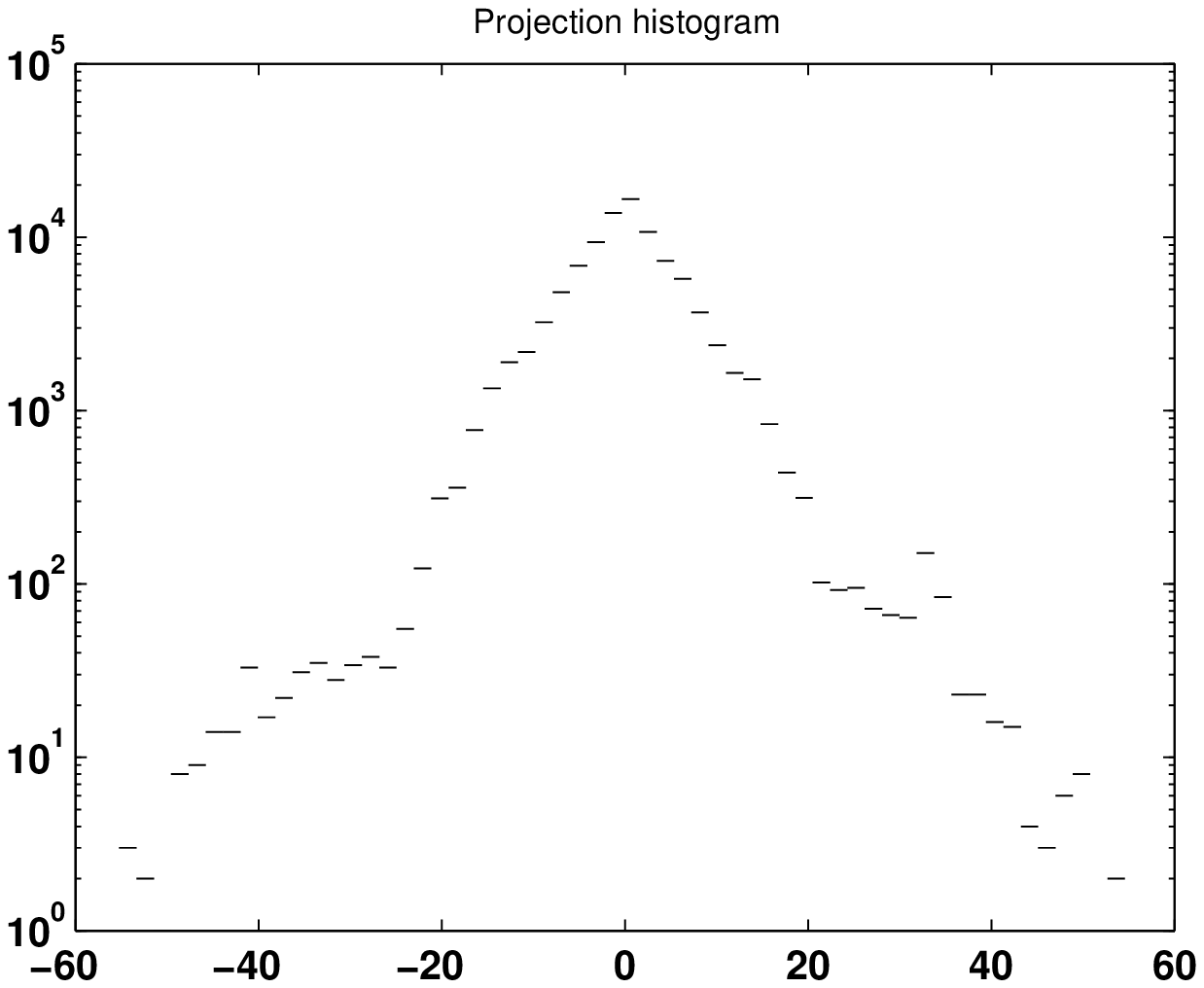}
 \end{center}
 \caption{The distribution of the projection value for square shape with a size 48x48 pixels.
  }\label{figdist}
\end{minipage}
\end{center}
\end{figure}

As mentioned above we must look for outliers in the distribution.
This would be difficult in the case of many 
dimensions, but easier in the case of one dimensional projection. 

We will regard only projections orthogonal to the vector with components 
proportional to $X_0(\rho)=1,\forall \rho$.
The projection on the direction of $X_0$ is proportional to the mean 
brightness of the area and thus can be considered as not so important 
characteristics of the image. An alternative interpretation of the above
statement is by considering all blocks that differ 
only by their brightness to be equivalent.

Mathematically the projections orthogonal to $X_0$ have the property:
\begin{equation}
\sum_{\vec \rho} X(\vec \rho)=0.
\label{projpr}
\end{equation}

The distribution of the values of the projections satisfying the 
property Eq.(\ref{projpr}) is well known and universal \cite{rudin} for the 2D natural images. 
The same distribution seems to be valid for a vast majority of the images. 
The distribution of the projections derived for the X-ray image, shown in
Fig. \ref{figorig}, is shown in Fig. \ref{figdist}.

\begin{figure}[t]
\begin{center}
\begin{minipage}{0.45\textwidth}
 \epsfysize=3.9cm
 \begin{center}
 \epsfbox{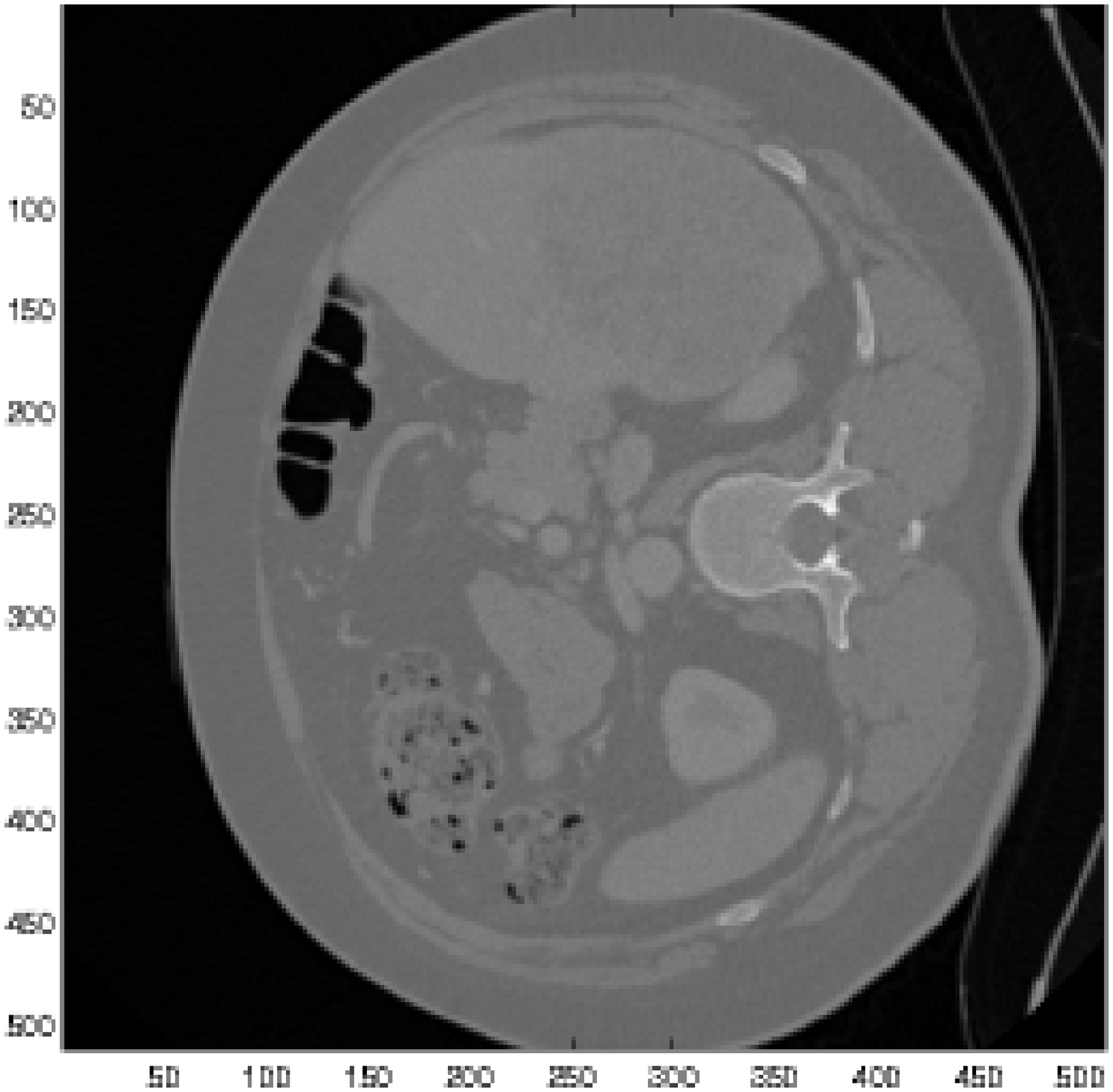}
 \end{center}
 \caption{An intersection of the 3d-test image. 
 }\label{section_48}
\end{minipage}
\hfill
\begin{minipage}{0.45\textwidth}
 \epsfysize 4.2cm 
 \begin{center}
 \epsfbox{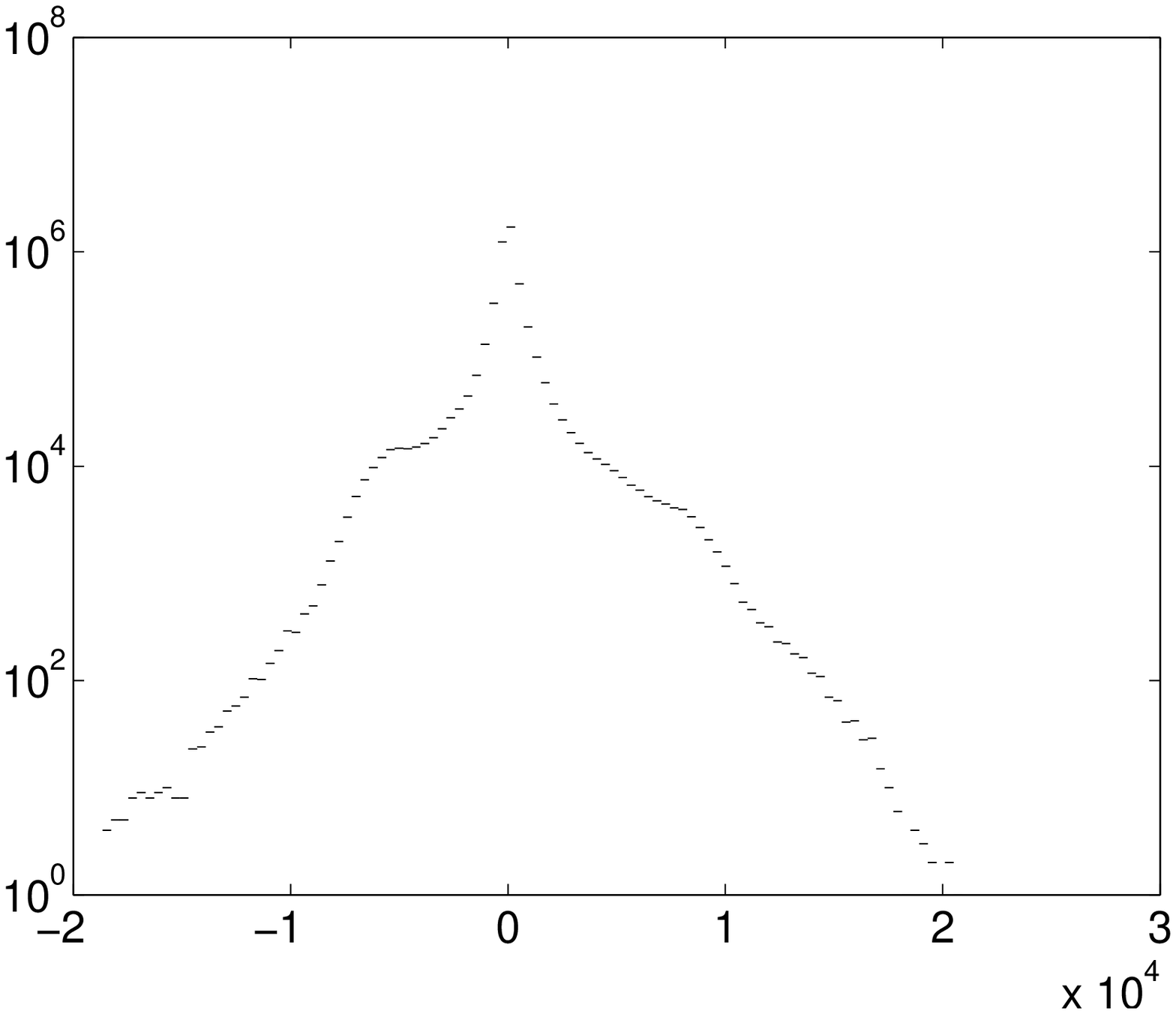}
 \end{center}
 \caption{The distribution of the projection value for the
   corresponding 3d image.
  }\label{hist3d_16_16_16}
\end{minipage}
\end{center}
\end{figure}

In the case of a three-dimensional image, Fig. \ref{section_48}, the corresponding histogram, obtained by using the
  above method is shown in Fig. \ref{hist3d_16_16_16}. 
 One observes a higher asymmetry of the distribution of the
 projections in this case, compared to the same distribution of two-dimensional images, 
{  but qualitatively it is of the same type.}

{  Roughly speaking, if the blocks are small enough,
the distribution satisfies a power law distribution 
with exponential drop at the extremes.}
When the blocks are big enough, the exponential part is predominant.

If $A_r$ and $A_r'$ have similar projections, then they 
will belong to one and the same or to adjacent bins. 
Therefore we can look for blocks that have a minimal number of 
similar and large projections. But these, due to the universality of the 
distribution, are exactly the blocks with large projection values.

As a first approximation, we can just consider the projections and score the 
points according to the bin they belong to. The distribution can be 
described by only one parameter that, for convenience, can be chosen to be the 
standard deviation $\sigma_X$ of the distribution of $X.B$.

The notion of ''large value
of the projection'' will be different for different projections but will be 
always proportional to the standard deviation\footnote{In general, the
standard 
deviation will be larger for projections with larger low-frequency
components. That is why we choose the criterion proportional to $\sigma_X$ and 
not as an absolute value {  for all projections $X$}.}.
Therefore we can define a parameter $a$ and score the blocks 
with $|X.B|>a \sigma_X$.

\subsection{Algorithm}

{  Resuming, 
in order to find the most unusual blocks of shape $S$ in an image $A$, we propose the following}

{\bf  Algorithm:}

\vskip2mm
0. Initialize: Construct a figure $B$ with the same shape as $A$ and with all pixels equal to zero. The result of the algorithm will be saved in $B$.

1. Generate a random projection operator $X$, with carrier 
with shape $S$, zero mean and norm one.

2. Project all blocks (convolute the figure). We denote 
the resulting figure as $C$.

3. Calculate the standard derivation $\sigma_X$ of the result of the convolution.

4. For all points of $C$ with absolute values greater than $a \sigma_X$, 
increment the corresponding pixel in B.

Repeat steps 1-4 for $M$ number of times.

5. Select the maximal values of $B$ as the most singular part of the image.

{  The acceptable values of $a$ are discussed in the next section.} The number of iterations $M$ can be fixed empirically or until the changes in $B$, 
normalized by that number, become insignificant.
Following the algorithm, one can see that the time to perform it 
is proportional to $M N^d \log N$. The speed per image of size 
$1024\times 2048$ on one and the same computer, 
with $S$, a square of size $56\times 56$ points,
is about 3 seconds compared to about an hour, using the direct search by implementing Eq. (1). We use a laptop with 
2GHz Intel Celeron CPU and 1GB of memory.
\footnote{If the block is small enough, the convolution 
can be performed even faster in the space domain and it is possible to 
improve the execution time.}

\begin{figure}[t]
 \begin{center}
  \begin{minipage}{2.90cm}
    \epsfxsize=2.8cm
    \epsfbox{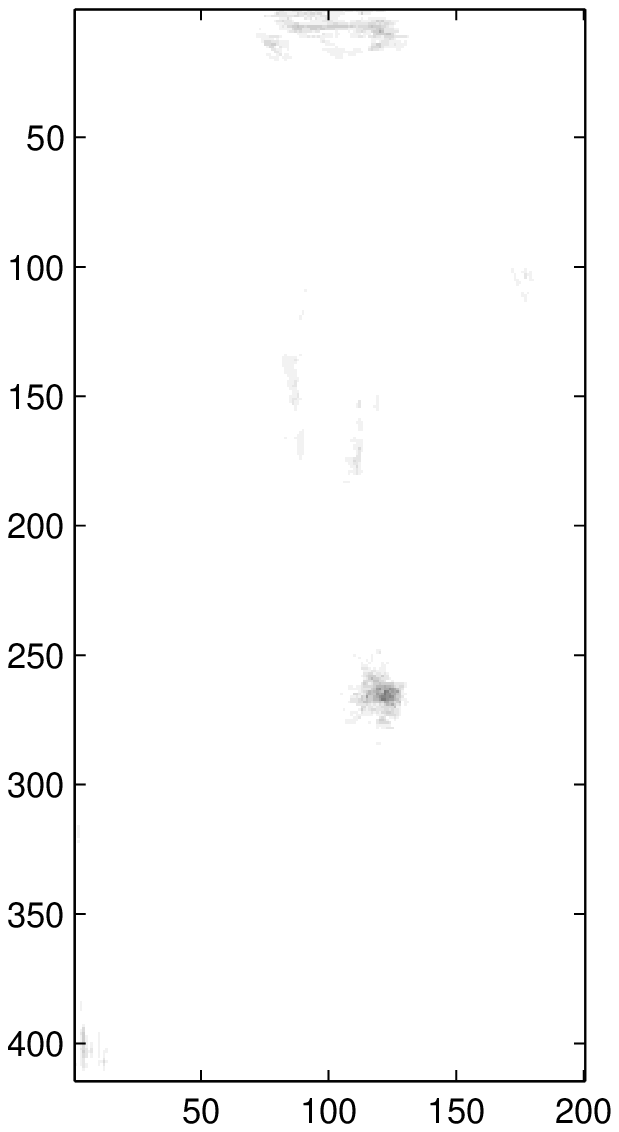}\\24x24
  \end{minipage}
  \begin{minipage}{2.90cm}
    \epsfxsize=2.8cm
    \epsfbox{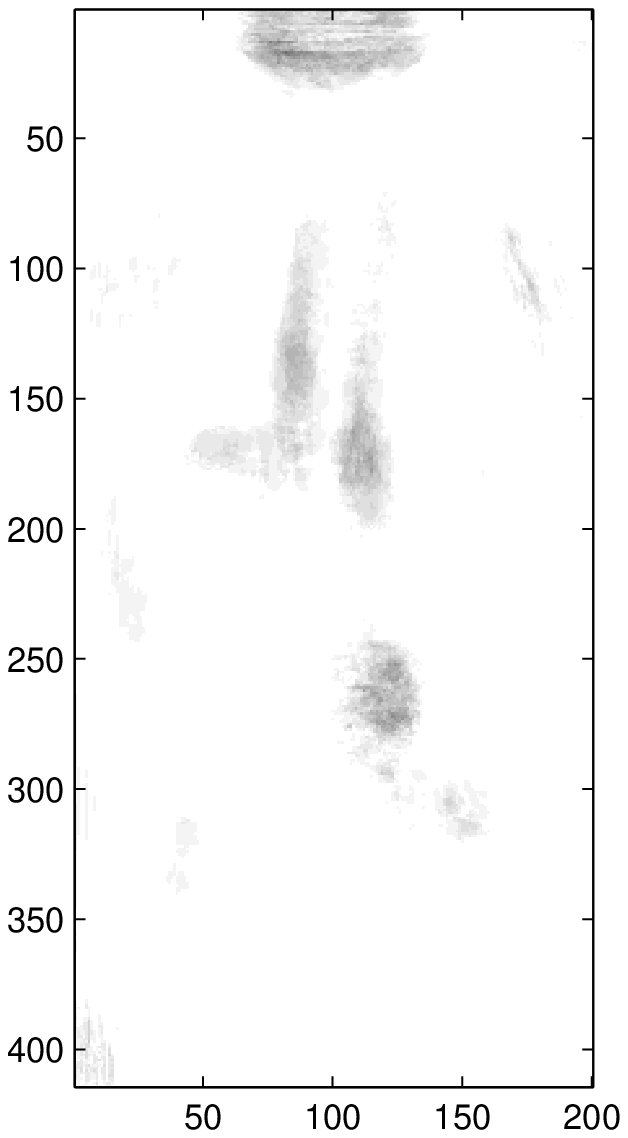}\\32x32
  \end{minipage}
  \begin{minipage}{2.90cm}
    \epsfxsize=2.8cm
    \epsfbox{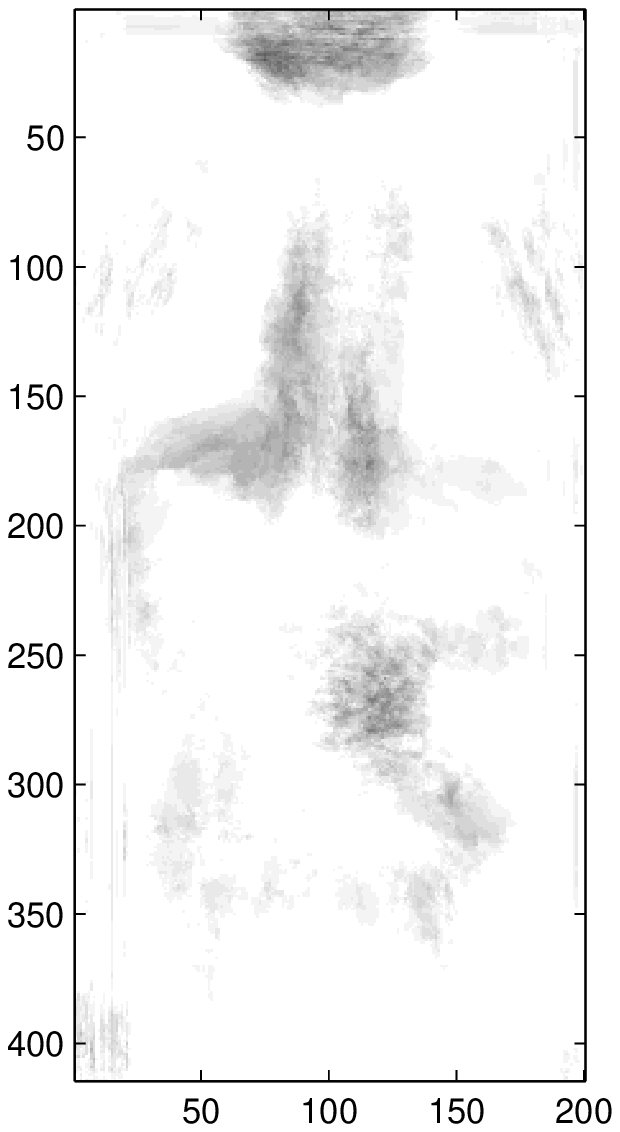}\\40x40
  \end{minipage}
  \begin{minipage}{2.90cm}
    \epsfxsize=2.8cm
    \epsfbox{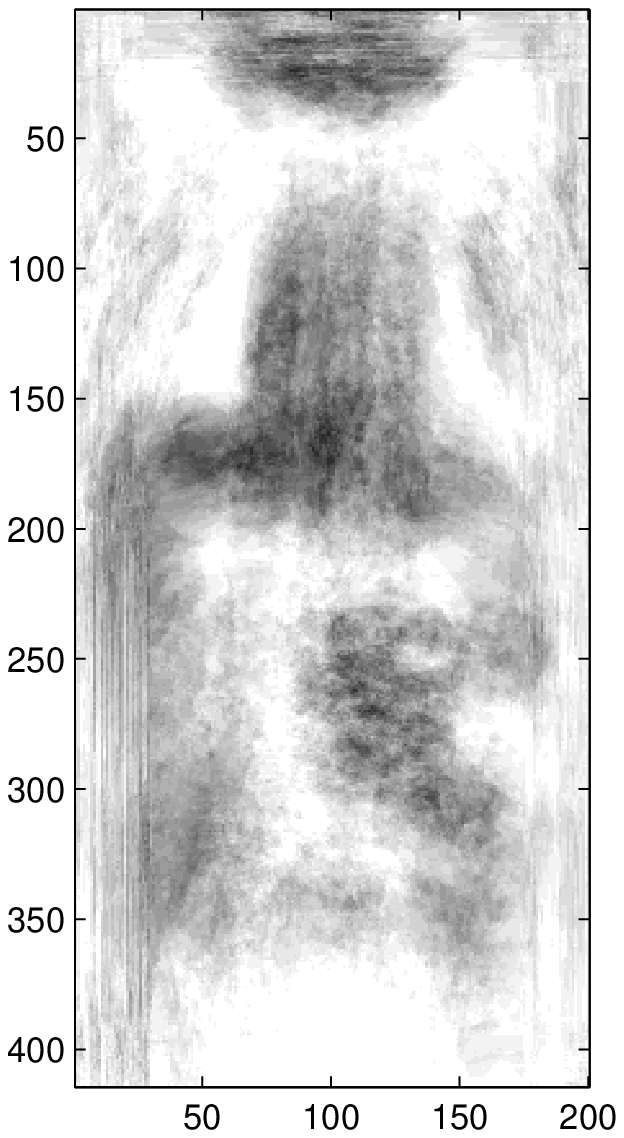}\\56x56
  \end{minipage}
 \end{center}
 \caption{Score values for different size of the shape
 (24x24, 32x32, 40x40, 56x56). 
The value of the parameter $a$ in all the cases is 12.}\label{mainfig}
\end{figure}
\begin{figure}[t]
\begin{center}
\begin{minipage}{2.90cm}
\epsfxsize=2.8cm
 \epsfbox{figure3.eps}\\$a=8$
\end{minipage}
\begin{minipage}{2.90cm}
\epsfxsize=2.8cm
 \epsfbox{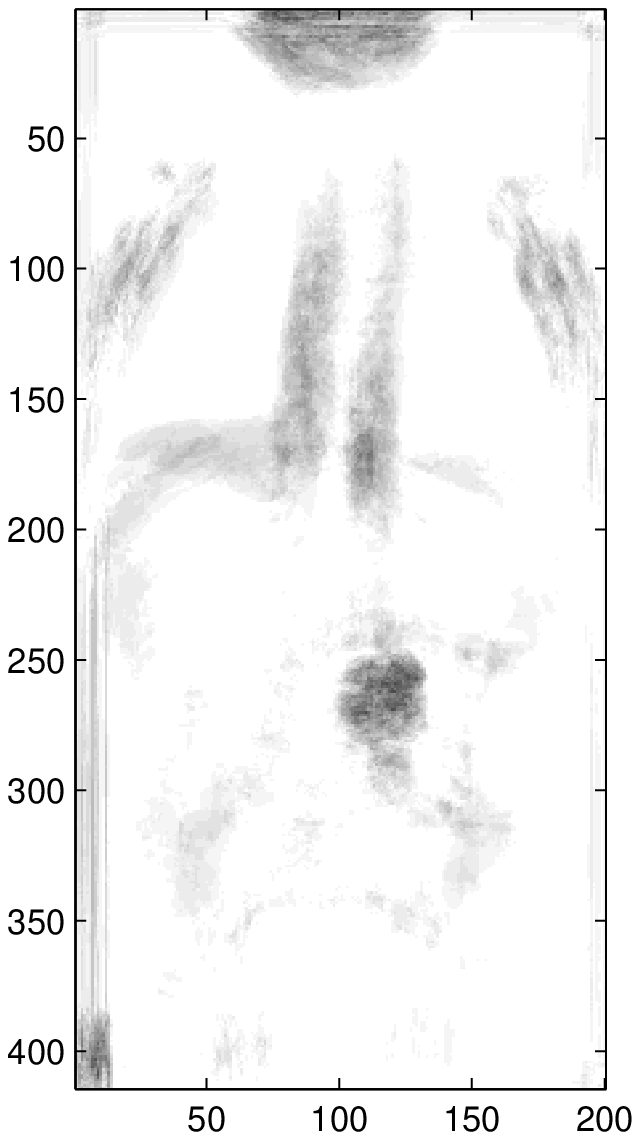}\\$a=10$
\end{minipage}
\begin{minipage}{2.90cm}
\epsfxsize=2.8cm
 \epsfbox{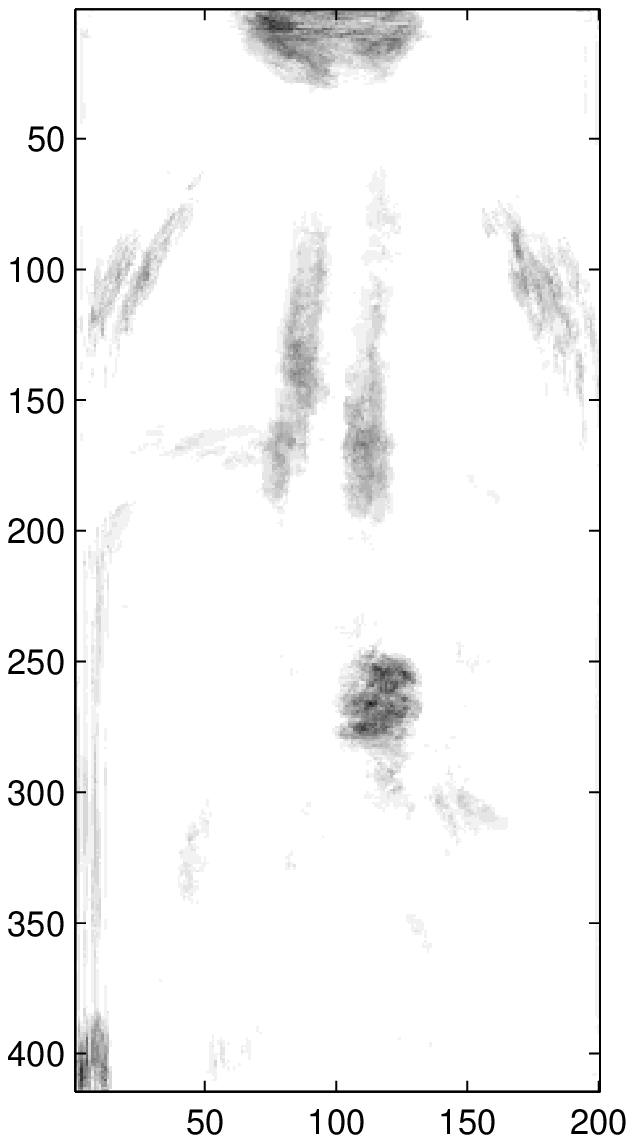}\\$a=12$
\end{minipage}
\begin{minipage}{2.90cm}
\epsfxsize=2.8cm
 \epsfbox{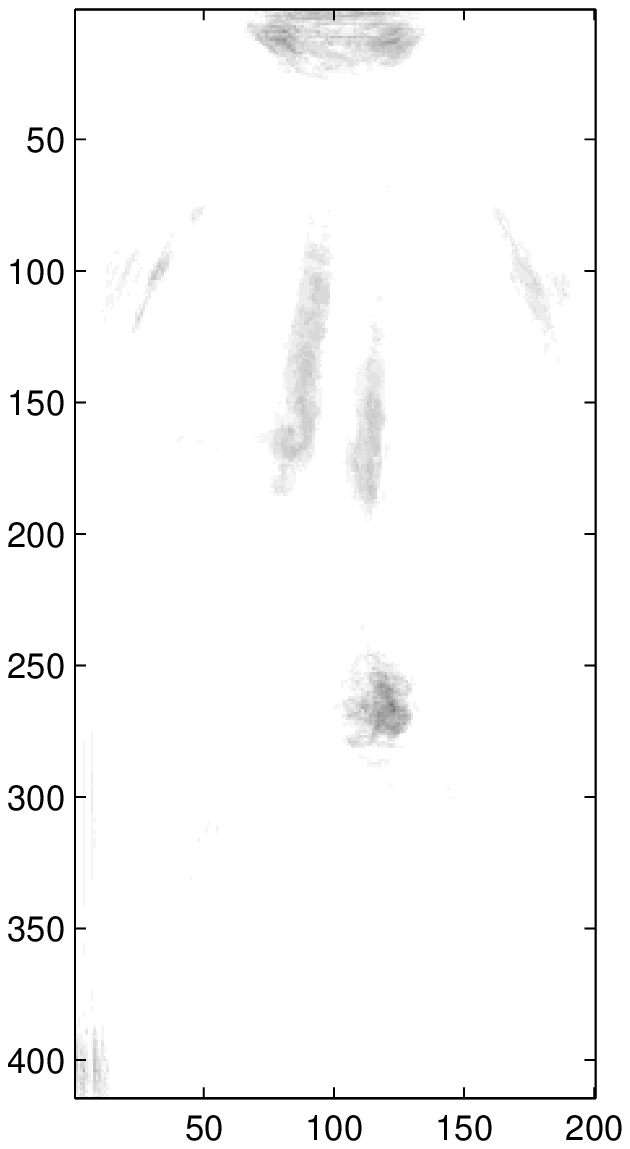}\\$a=16$
\end{minipage}
\end{center}
 \caption{Score values for different parameter $a$ ($a$ 
= 8,10,12,16). The size of the shape is 24x24.
 }\label{mainfig1}
\end{figure}

\section{Empirical Assessment}

Applying the algorithm above, we are looking for the 
most unusual part of the image in different settings. 
We also generalize the method in order to improve it 
and in to amplify the range of applications.

\subsection{Two dimensional images}
Some results are presented in Figs.~\ref{mainfig},\ref{mainfig1}, where we used square shapes with 
different size, 30 random projection operators and different values of $a$.

Because the distribution of the projections (Fig. 2) is universal,
 it is not surprising that the algorithm is operational for different images. 
We have tested it with some 100 medical Xray images and the results of the 
visual inspections were good
.

It can be noted that the number of projection operators is not critical and can be kept relatively low and independent of the size of the block. Note that with significantly large blocks, the results can not be regarded as an edge detector. This empirical observation is not a trivial result at all, indicating that the degrees of freedom are relatively few, even with large enough blocks, something that depends on the statistics of the images and can not be stated in general. With more than 20 projections we achieve satisfactory results, even for areas with more than 3000 pixels {  (some $10^5$ in 3D)}. The increment of the number of the projections improves the quality, but with more than 30 projection practically no improvement can be observed.

{  A phenomenological argument can be given, observing that in the case of 30 projections, the pixels with maximal values are larger than 5. In order to distinguish a binary criteria (unusual/usual) this value is satisfactory large.}

It is possible to look at that algorithm in a different way, namely, if we 
are trying to reconstruct the figure by using some projection 
operators $X_C$ (for example DCT as in JPEG), 
then the length of the code, one uses to code a component with distribution 
like 
Fig. \ref{figdist}, will be proportional to the logarithm of the 
probability of some value of the projection $X_C.A$.
Therefore, what we are scoring is the block that has some component of the 
code larger than some length in bits (here we ignore the psychometric 
aspects of the coding). 
Effectively we score the blocks with longer coding, e.g. the 
ones that have lower probability of occurrence.

Using a smoothed version of the above algorithm in step 4, without 
adding only one or zero,
but for example, penalizing the 
point with the square of the projection difference with respect to the current 
block divided by $\sigma$, and having in mind the universal distribution of the 
projection, one can compute the penalty function as a function of the value of 
the projection $x$, that results to be just $1/2+x^2/2\sigma^2$. Summing over 
all projections, we can obtain that the probability of finding the best block is 
approximated given by 
$1/2[1+{\rm erfc}(M(1/2+x^2/2\sigma^2))]$ as a consequence of the Central 
Limit Theorem. {The above estimation gives an idea why one needs few projections to find the most unusual block, 
in sense of the global distribution of the blocks,
almost independently of the
size 
of the block.} The only dependence of the size of the blocks is given by 
$\sigma^2$ factor, that is proportional to its size. 
{Further, the probability of error will drop better than exponentially with 
the increment of $M$.}

The non-smoothed version performs somewhat better that the above estimation
in the computer experiments. 

\subsection{Three dimensional images}

\begin{figure}[t]
\begin{center}
\begin{minipage}{5.60cm}
 \epsfxsize=5.2cm
 \epsfbox{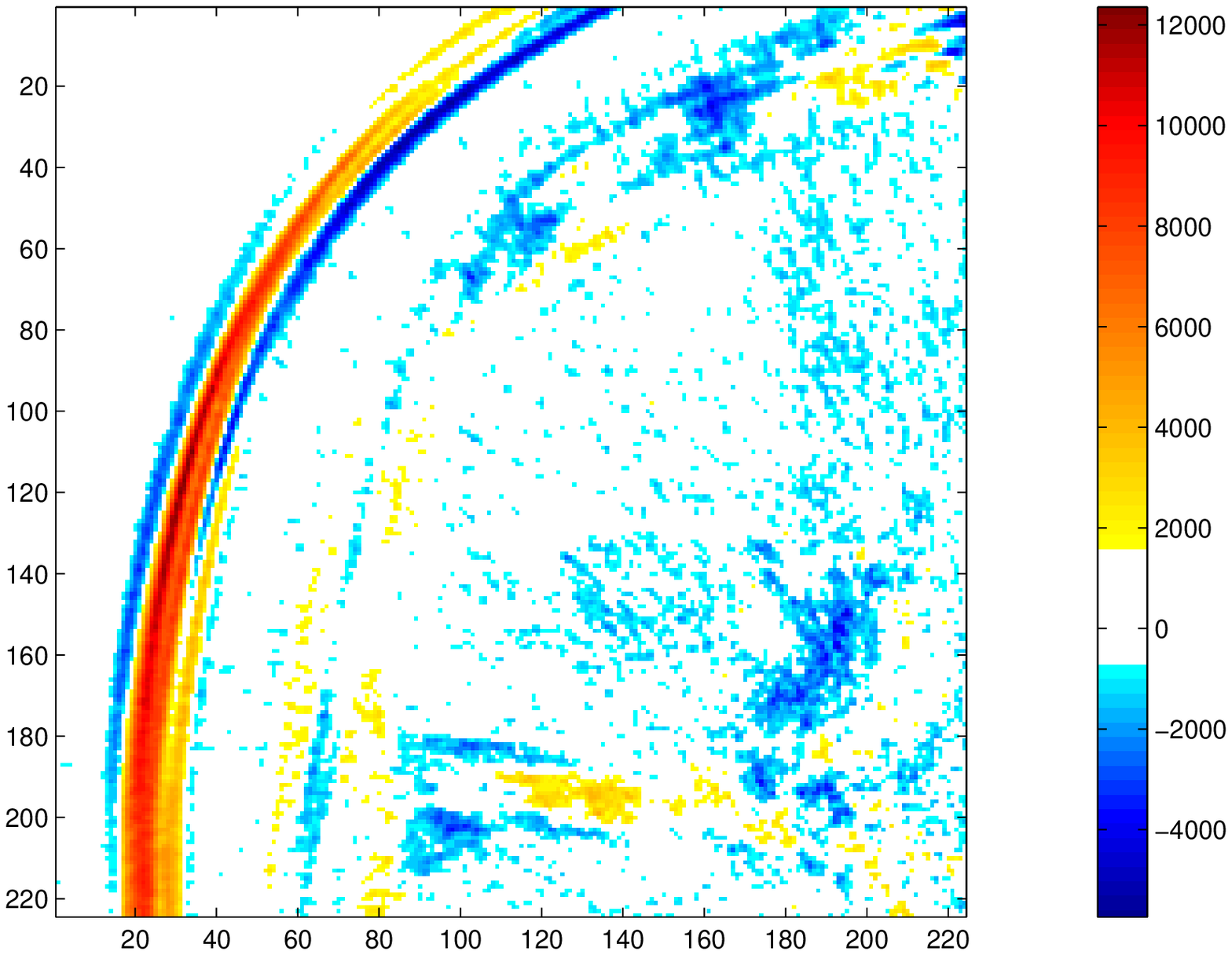}
\end{minipage}
\begin{minipage}{5.60cm}
 \epsfxsize=5.2cm
 \epsfbox{ 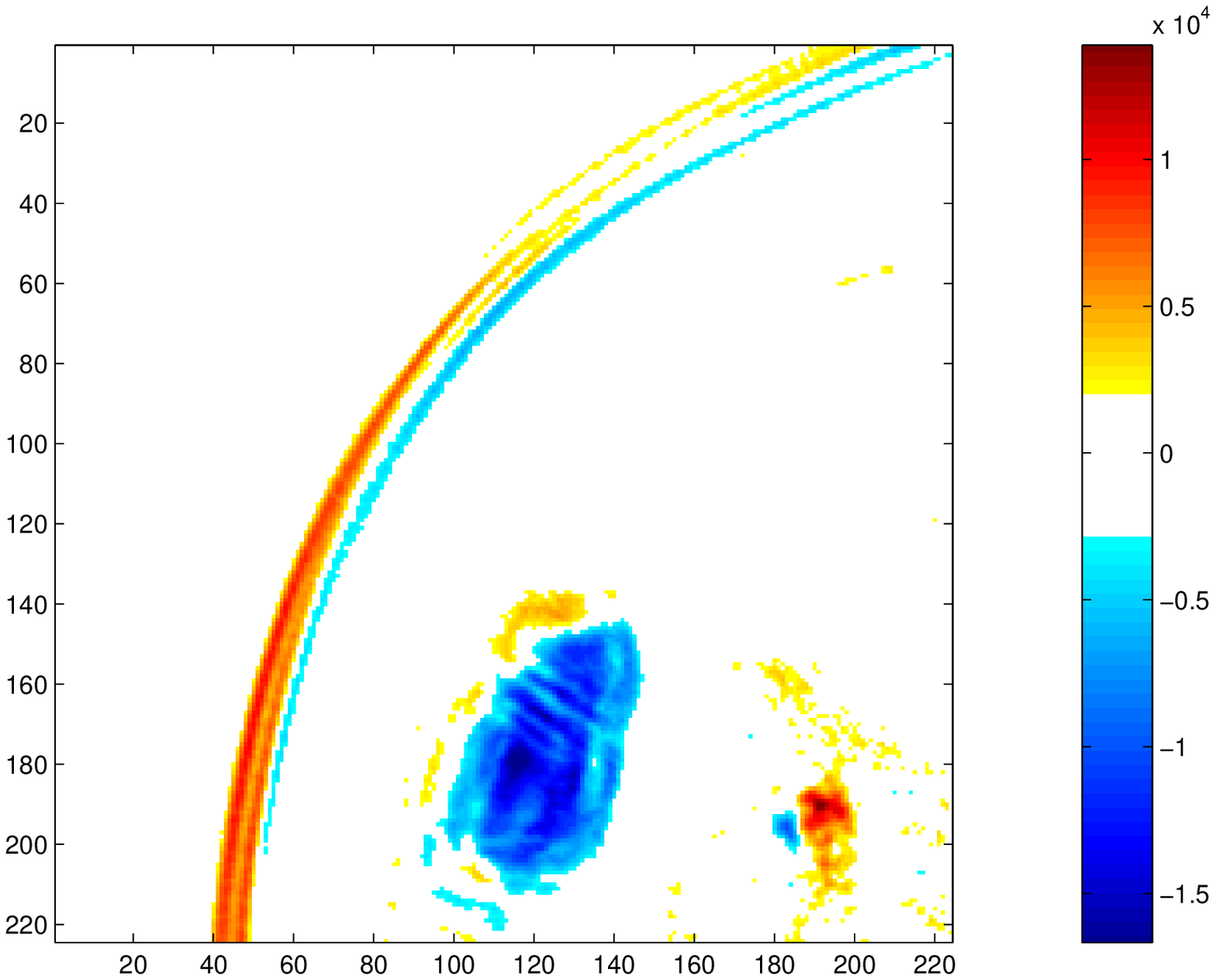}
\end{minipage}
\caption{
The upper panel shows the 3D structure, treated with 3D shapes.
The lower panel shows the same structure treated with 2D shapes (at the same value of Z axes). The size of the block is $24\times 24$ and $24\times 24\times 24$ correspondingly.
The figures show that if the structures are clearly 3D the detection with 3D shape is better.
}\label{fig2d3d}
\end{center}
\end{figure}

Further we investigate how the algorithm works in 3D. As noticed in the previous section, the distribution of the projection is similar, but more irregular and asymmetric. 

{ We have noticed that in 3D the method works better by using spherical shape, instead of cubic one.
Most probably this is due to the fact that in the cube the most distant boundary 
voxel is $\sqrt{3}$ times further that the closest boundary voxel. This is significantly more that $\sqrt{2}$ as it is in the case of 2D pixels, although some "squaring" effect can be noted also in 2D (See Fig.\ref{gammas}, $\gamma=0.5$).}

Comparing the quality of the method on two and three-dimensional
images, (Fig. \ref{fig2d3d}), 
one can say that when the structure is clearly three dimensional, the algorithm working in 3D separates this structure much better than working in 2D section.
The irrelevance of the dimension for the algorithm is probably the main advantage with respect to other algorithms, as for example the Hough transform \cite{Hough}. 
The maximum execution time scales with $N$ as  
the number of pixels $M N^d \log_2 N^d$. 
Once again in 3D, as in the 2D case, 
$M$ can be chosen very modest, about 30. 
The memory cost is four times the memory 
needed to save a single image, 
using naive FFT implementation of the convolution.

\subsection{Contrast}

\begin{figure}[t]
\begin{center}
\begin{minipage}{4.20cm}
 \epsfxsize=4.0cm
 \epsfbox{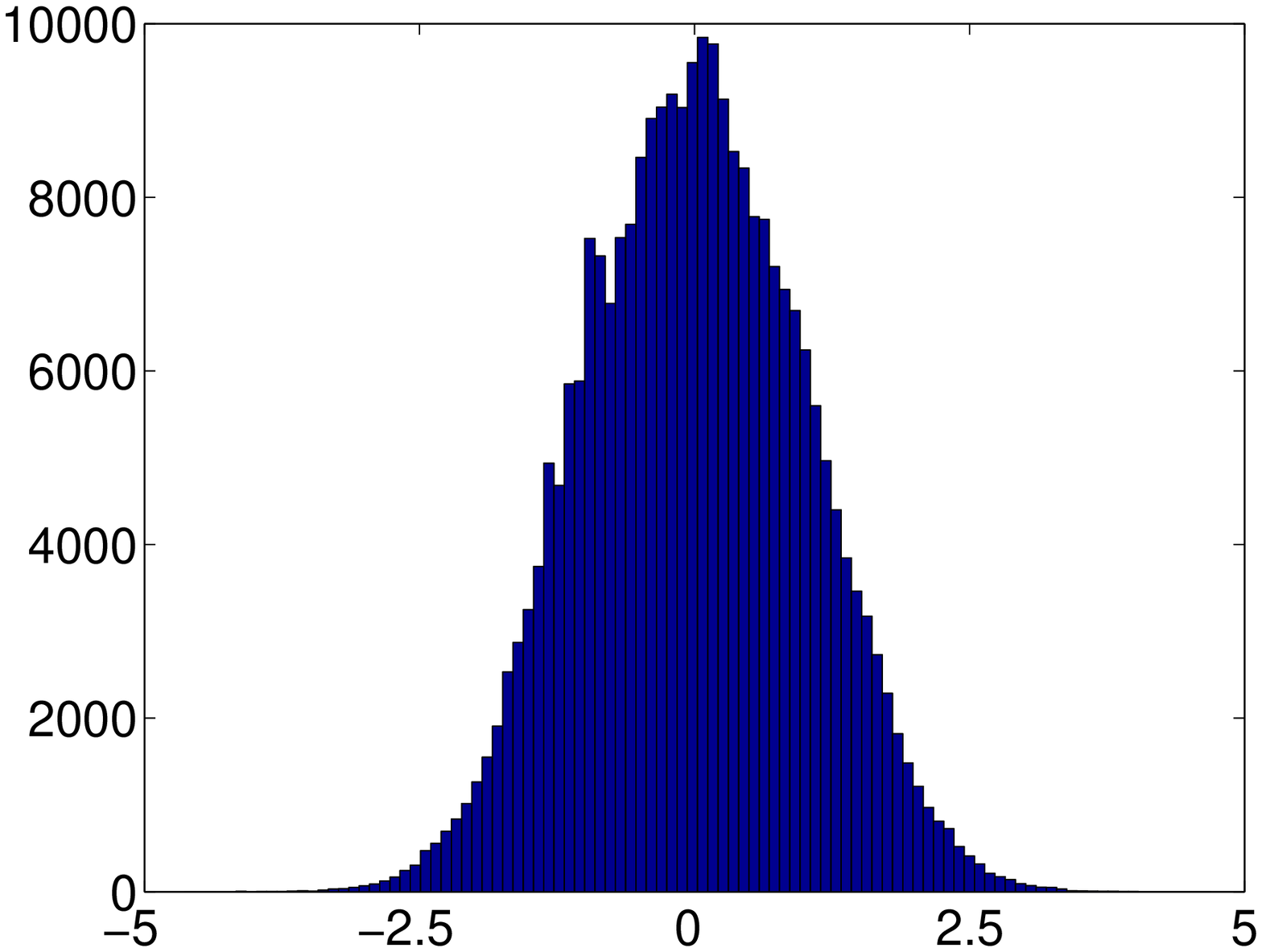}
\end{minipage}
\begin{minipage}{4.20cm}
 \epsfxsize=4.0cm
 \epsfbox{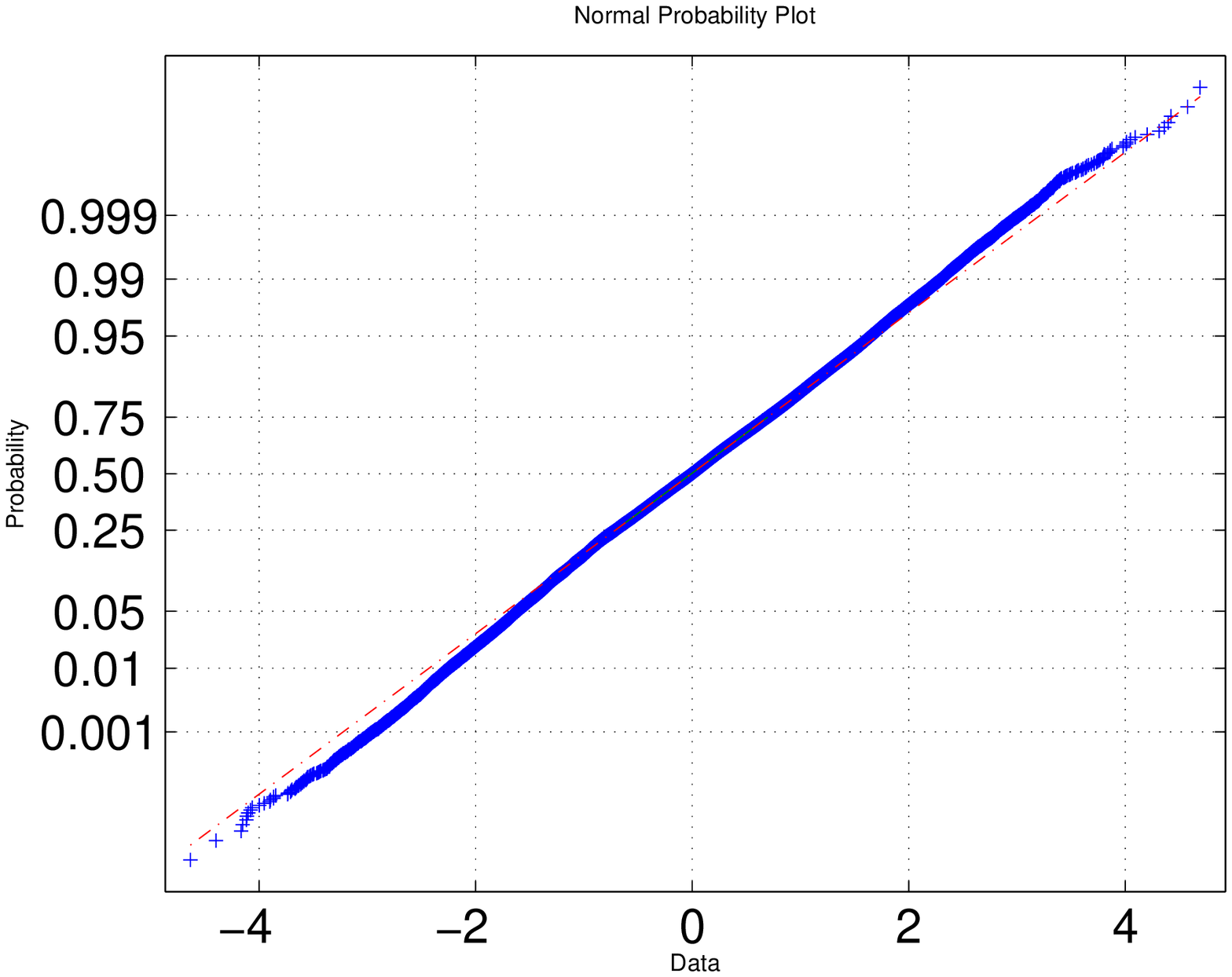}
\end{minipage}
\caption{
The left panel shows the histogram of the normalized projection $X_B'=X_B/\sigma_B$. The right one shows the quintile norm-plot. The distribution fits very well with the normal distribution.
}\label{quadnorm}
\end{center}
\end{figure}

However, there is an evident objection against the proposed algorithm. Namely, if some area of the image with high contrast is selected, then the projection is proportional to the contrast of that area. This will actually select the most contrast areas as the most unusual ones. Mathematically this is in fact so, the boundaries are the most unusual parts of the images, but for practical reasons the dependence of the contrast should be eliminated or at least attenuated with similar argumentation as the one we have used for the brightness. 

Namely, two images that differ only by their contrast could be considered as equivalent. To eliminate the influence of the contrast, the best is to normalize the projections using the contrast of the block. Let us regard as a contrast the standard deviation $\sigma_B$ of the block $B$ in question. Then the change in the algorithm is just evident: 
substitute each of the projections Eq.(\ref{projdef}) with its normalized value $X_B'\equiv X_B/\sigma_B$. 
However, the distribution $X_B'$ is no longer similar to that shown in Fig. \ref{hist3d_16_16_16}. 
The distribution is just normal \cite{kkdorron}. 
To illustrate this, we represent the distribution and its quartile normal-plot in Fig.\ref{quadnorm} \footnote{The distribution ought to be tested with caution because the low-pass filtering will flat the top of the distribution. Also the precision of the pixels ought to be at least 2 bytes in order to avoid rounding errors.}. 

\begin{figure*}[th]
\begin{center}
\begin{minipage}{5.60cm}
 \epsfxsize=5.0cm
 \epsfbox{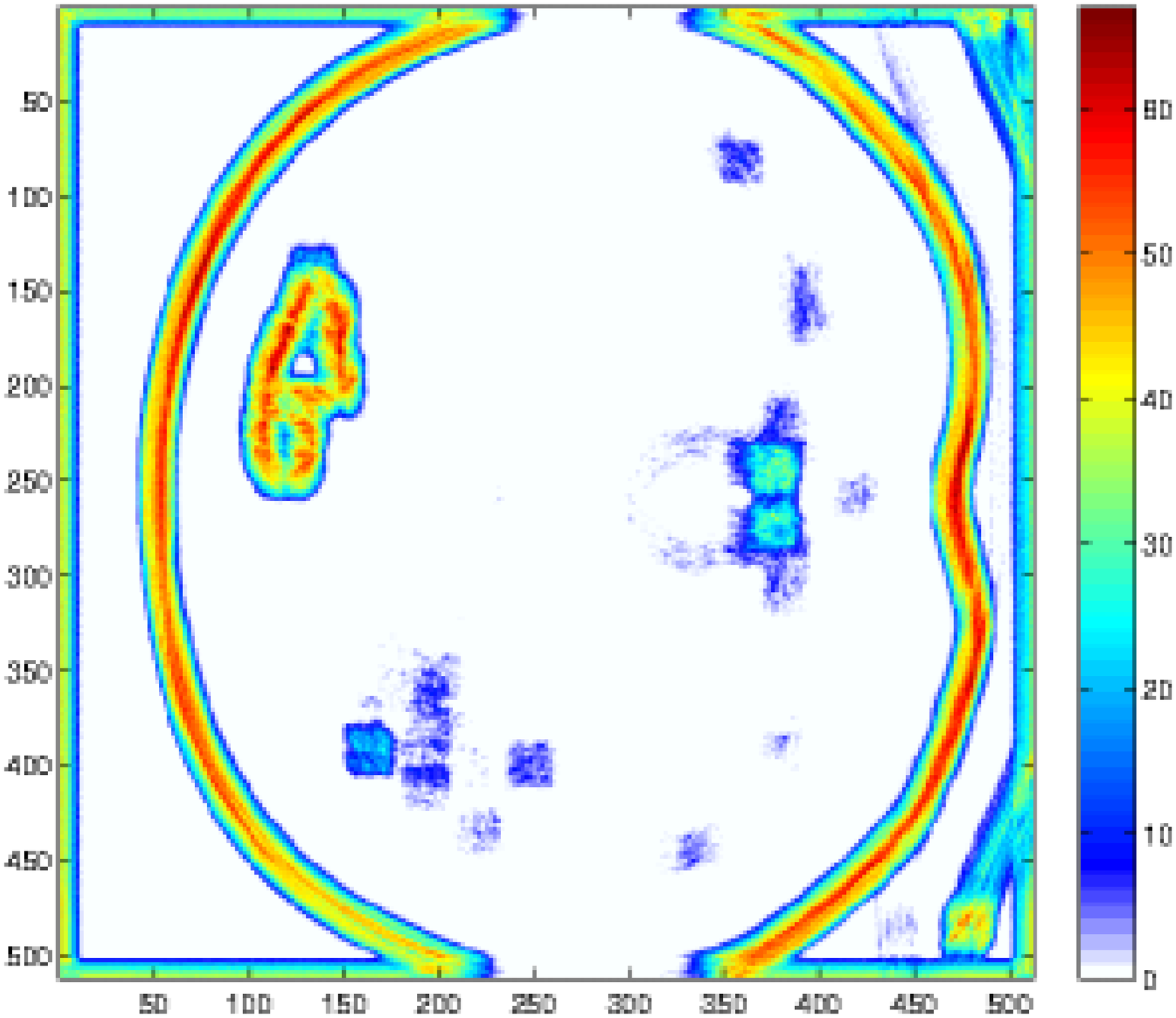}\\$\gamma=0$
\end{minipage}
\begin{minipage}{5.60cm}
 \epsfxsize=5.0cm
 \epsfbox{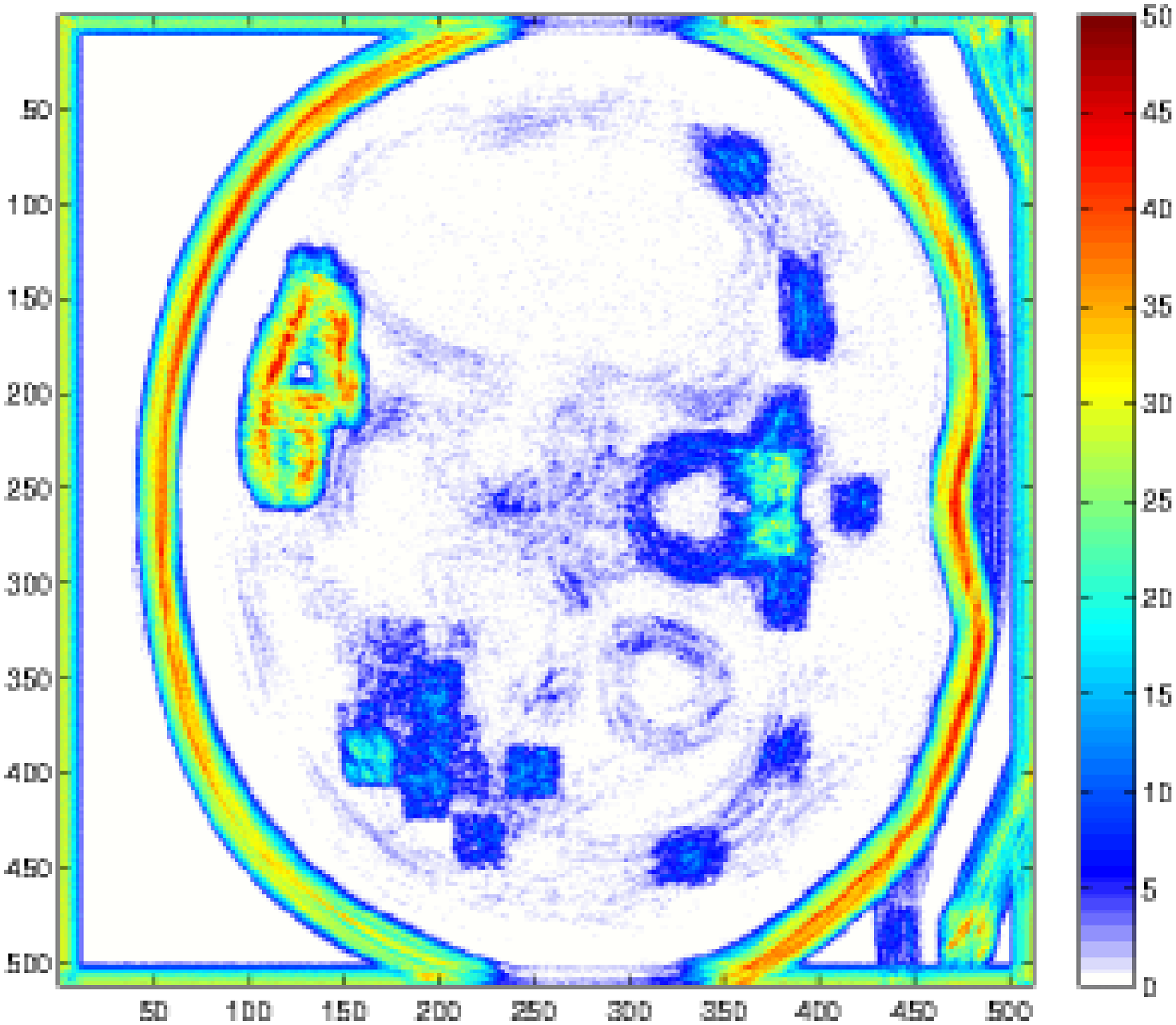}\\$\gamma=0.5$
\end{minipage}\\
%
\begin{minipage}{5.60cm}
 \epsfxsize=5.0cm
 \epsfbox{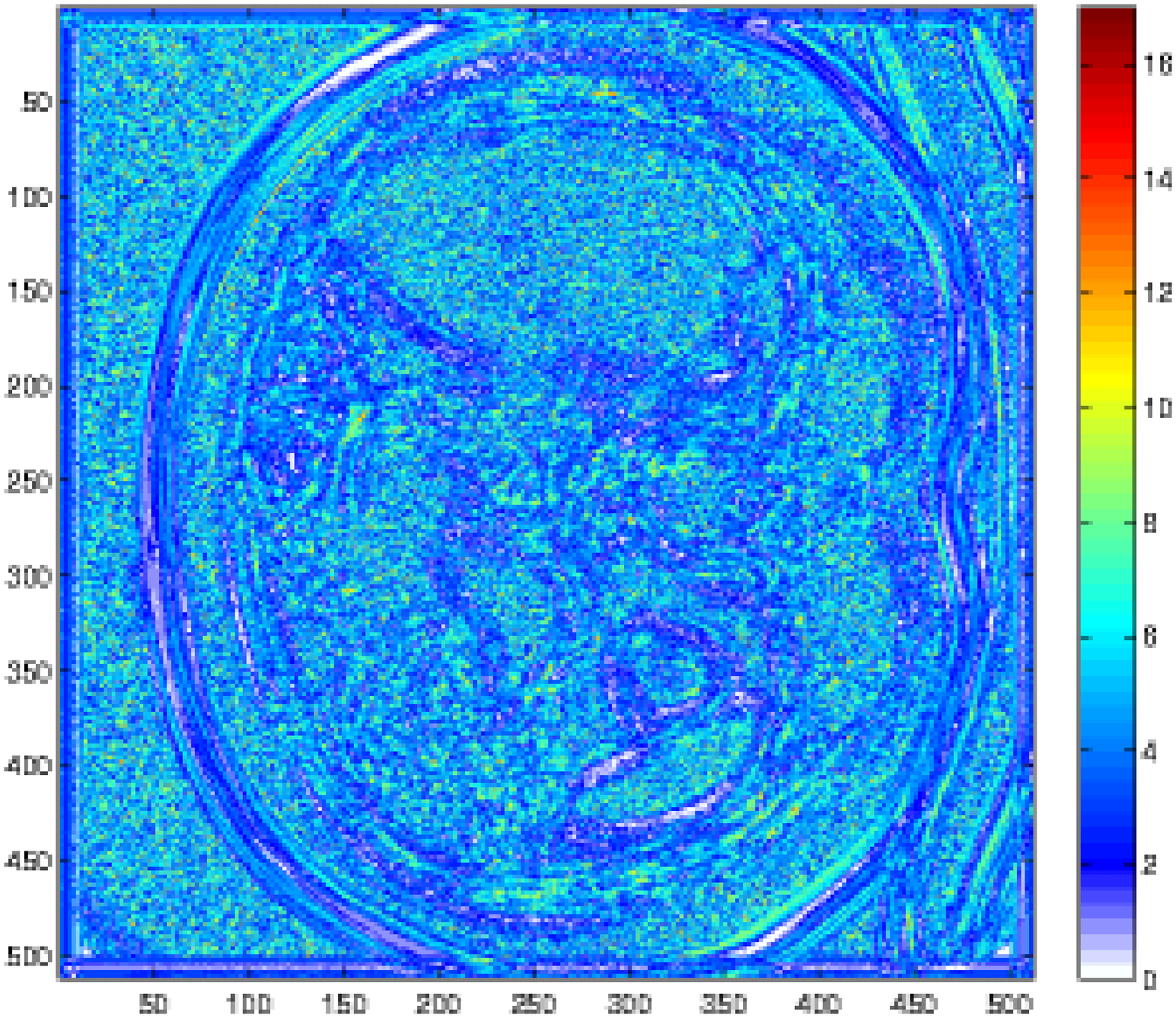}\\$\gamma=1$
\end{minipage}
\begin{minipage}{5.60cm}
 \epsfxsize=5.0cm
 \epsfbox{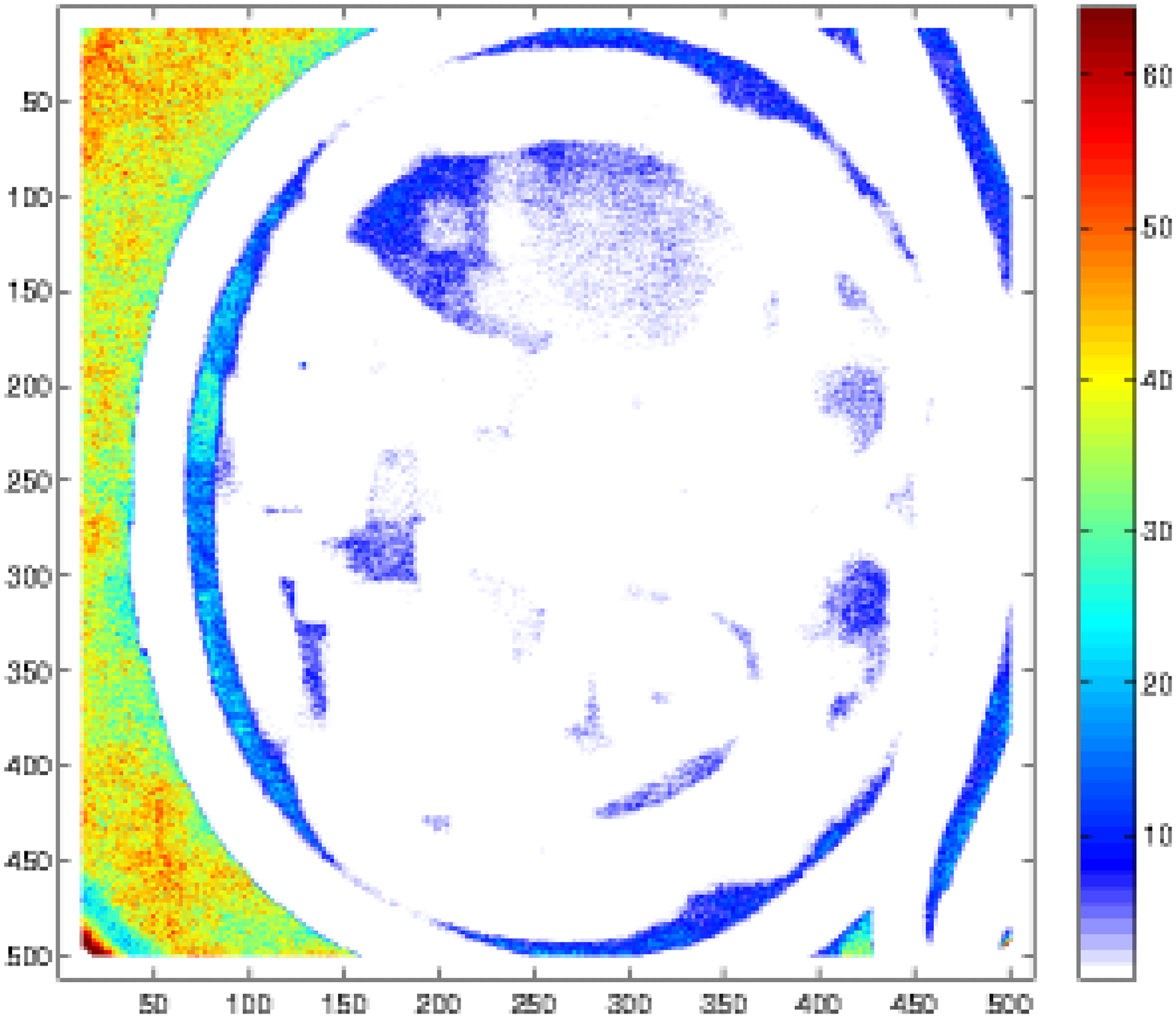}\\$\gamma=2$
\end{minipage}
 \caption{Different normalizations corresponding to the values of
   $\gamma=0, 0.5, 1, 2$. The influence of the borders diminishes and the influence of the texture increases.
 }\label{gammas}
\end{center}
\end{figure*}

 Using this normalization procedure makes the algorithm sensible to the noise, converting the flat noisy areas to the most unusual ones because of the randomness of the noise. 
Also the contrast, as an important characteristics, is better to be partially preserved in the normalized projection. Therefore it is much better not to eliminate the dependence of the contrast, but just to attenuate it. 
We found that using 
\[
X_B(\gamma)\equiv {X_B}/{\sigma_B^\gamma}
\]
with different $\gamma$-s serves well in order to give an appropriate weight of the contrast. When $\gamma=0$ we have the case { of uncorrected projections}, while when $\gamma=1$, the effect of the contrast is totally eliminated.  Also $\gamma$ can be assumed to be the tradeoff between the texture and the shape of the area.
{ In Fig.\ref{gammas} we represent the results for different $\gamma$-s ($\gamma=0, 0.5, 1, 1.5, 2$). 
We can see that different structures are highlighted dependent on the
values of gamma. 
As expected, low values of $\gamma$ (with relatively small shapes) accentuate the shape and
high values of $\gamma$ - the texture.}

\subsection{Network}

The pitfall of the consideration in the previous subsection is that the 
detected blocks are unusual in absolute sense, e.g. with respect to all figures
that 
satisfy the power law or similar distribution of the projections.
Actually this is not desirable. If for example, in X-ray image, several 
spinal segments appear, although these can be unusual in the context of all existing 
images, they are not unusual in the context of thorax or chest X-ray images.

Therefore the parts of the images with many similar projections 
must ``cancel'' each other.
This gives us the idea to build a network, where its components with similar 
projection are connected by a negative feedback
corresponding to
the blocks with similar projections.

As we have seen in the previous section, the small projection values are 
much more probable and therefore less informative. Using this empirical 
argument, we can suggest that the connections between the blocks with large 
projections are more significant.

The network is symmetrical by its nature, because of the reflexivity of the 
distances.
We can try to build it in a way similar to the Hebb network \cite{Hebb} 
and define 
Lyapunov o energy function of the network. Thus the network can be described in 
terms of artificial recursive neural network. 
Connecting only the elements of the image that produce large projections,
the network can be build extremely sparse \cite{sparse}, 
which makes it feasible in real cases.

Let us try to formalize the above considerations. 
For each point we define a  neuron.
The neurons corresponding to some point $\vec r$ and having projection $x$ 
receive a positive input flux, which is proportional to $-\log p(x) $, 
where $p$ is the probability of having projection with value $x$. 
The same element, if its projection is large, also receives a negative 
flux from the points $\vec r'$ with nearest projections 
that satisfy the condition
$|\vec r-\vec r'|>{\rm diam}(S)$.
The flux in general is a function of $p(x)$ and $x'-x$.

As a first 
approximation we assume that the flux is constant with $p(x)$ and the 
dependence on $x'-x$ is trivial:
the weight is 1 if $|x'-x|<\delta$ and zero otherwise, where $\delta$ is 
some parameter of the model.

In other words, we reformulate our problem in terms of a Hebb-like neural 
network with external field 
\begin{equation}
h=-h_0\sum_{i=1}^{M} \log p(x_i)
\end{equation}
 and weights 
\begin{equation}
w_{rr'}=-\sum_{i=1}^{M}\ \ \ \sum_{
\begin{array}{c}
|x_i|>a \sigma_i,\\ |x_i'|>a \sigma_i,\\ |x_{i}^{'}-x_i|<\delta,\ x_i' x_i >0
\end{array}
} 1.
\end{equation}
The extra parameter $h_0$ balances between the global and the local 
effects.
It can be chosen in a way that the mean fluxes of positive and negative
currents 
are equal in the whole network.
The parameter $\delta$, as a proof of concept value, can be assumed to be
equal to infinity.
So the only parameter, as in the previous case, is $a$.

The dynamics of the network over time $t$ is given by the following equation \cite{Hopfield}:
\[
s_{\vec r}(t+1)=g(\beta[h_r+\sum_{\vec r'} w_{\vec r\vec r'} s_{\vec r}(t)-T]),
\]
where $g(.)$ is a sigmoid function, $s_{\vec r}(t)$ is the state of the neuron $s$ at
position $\vec r$ and time $t$, $\beta$ is the inverse temperature  and 
$T$ is the threshold of the system. The result must be insensitive to the 
particular chose of $g(.)$.

\begin{figure}[t]
\begin{center}
\begin{minipage}{2.90cm}
 \epsfxsize=2.5cm  \epsfbox{figure10.eps}
\end{minipage}
\begin{minipage}{2.90cm}
 \epsfxsize=2.5cm
 \epsfbox{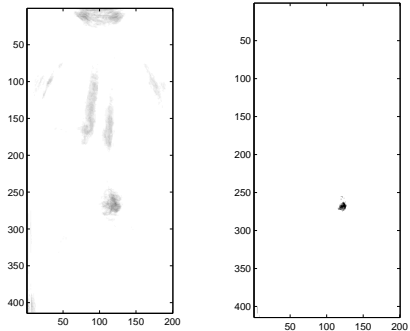}
\end{minipage}
 \caption{Comparison between score image (left) and network  activity image (right). 
The size of the area is 24x24 and the parameter $a=16$.
 }\label{figbump}
\end{center}
\end{figure}

Once the network is constructed, we need to choose its initial state. If the 
a priori probabilities for all points to be the origin of the most unusual block 
are 
equal, one can choose $s_{\vec r}(0)=1,\ \ \forall \vec r$.
Due to the non-linearity, the analysis of the results is not 
straightforward.
The existence of the attractor is guaranteed by the symmetrical nature of  
the
weights $w$, which is a necessary condition for the existence 
of an energy function.

\begin{figure*}[t!]
\begin{center}
\begin{minipage}{5.60cm}
 \epsfxsize=5.5cm  
 \epsfbox{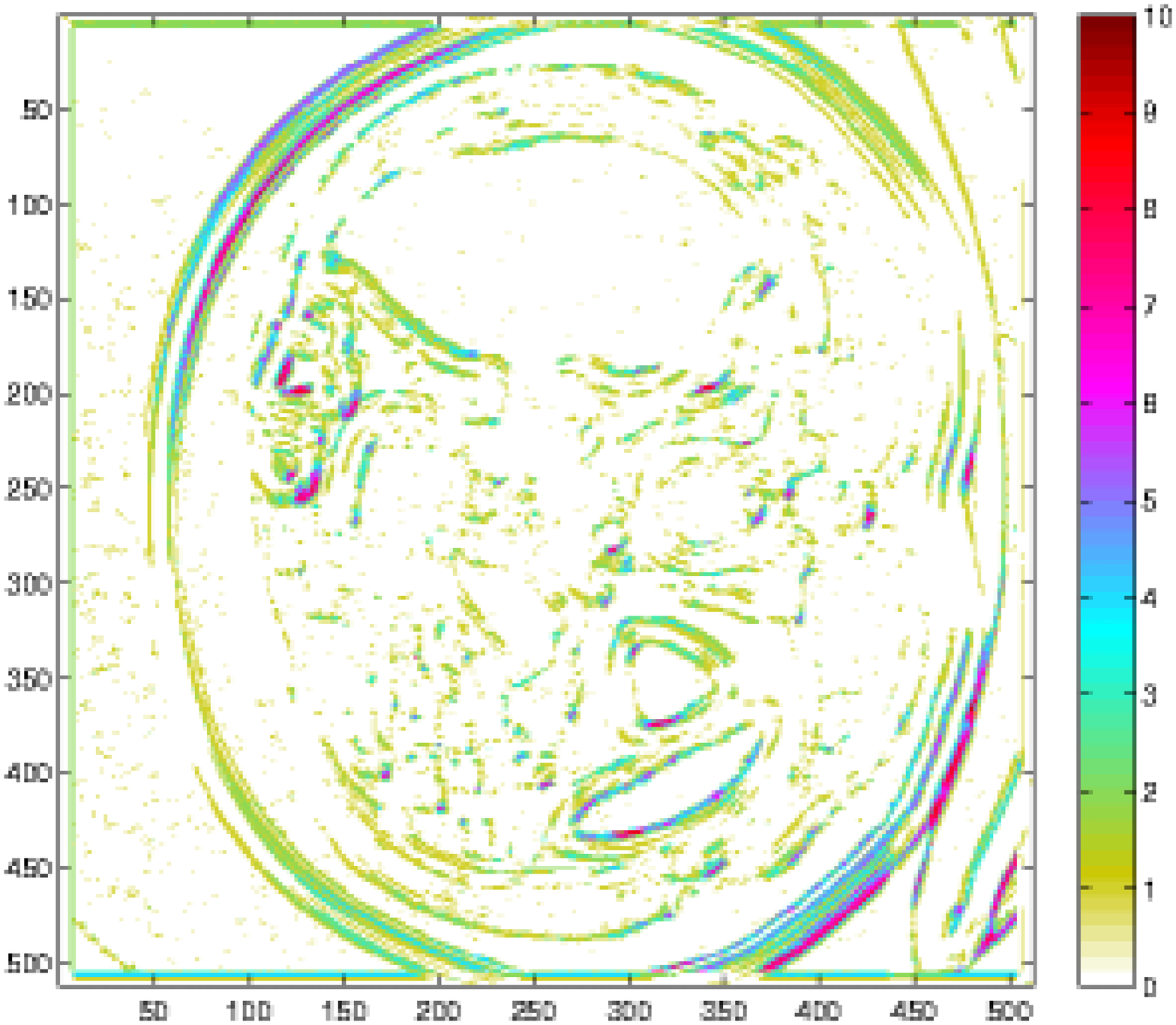}\\$16\times 16$
\end{minipage}
\begin{minipage}{5.60cm}
 \epsfxsize=5.5cm
 \vskip0.3cm 
  \epsfbox{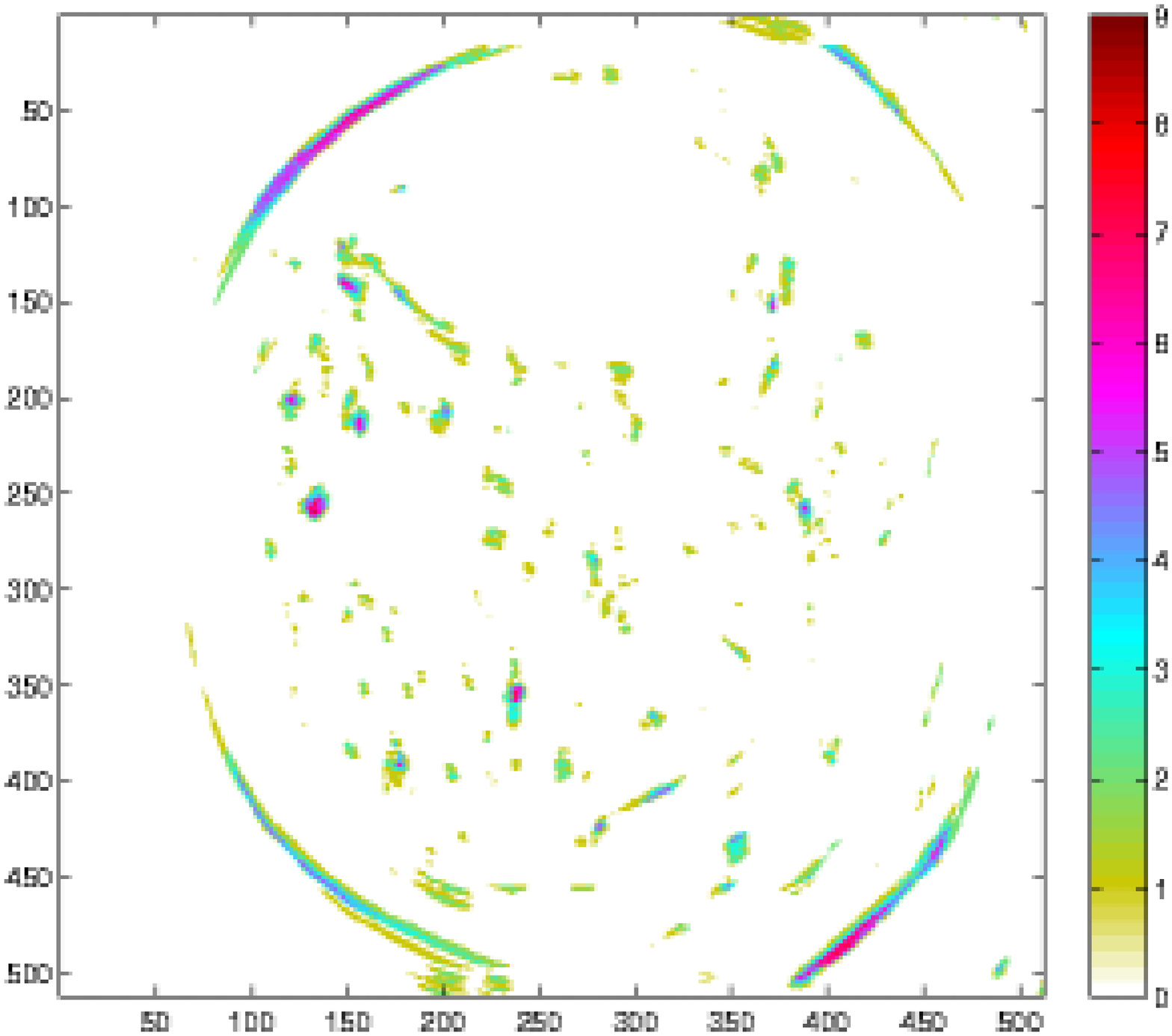}\\$32\times 32$
\end{minipage}\\
\begin{minipage}{5.60cm}
 \epsfxsize=5.5cm  
 \epsfbox{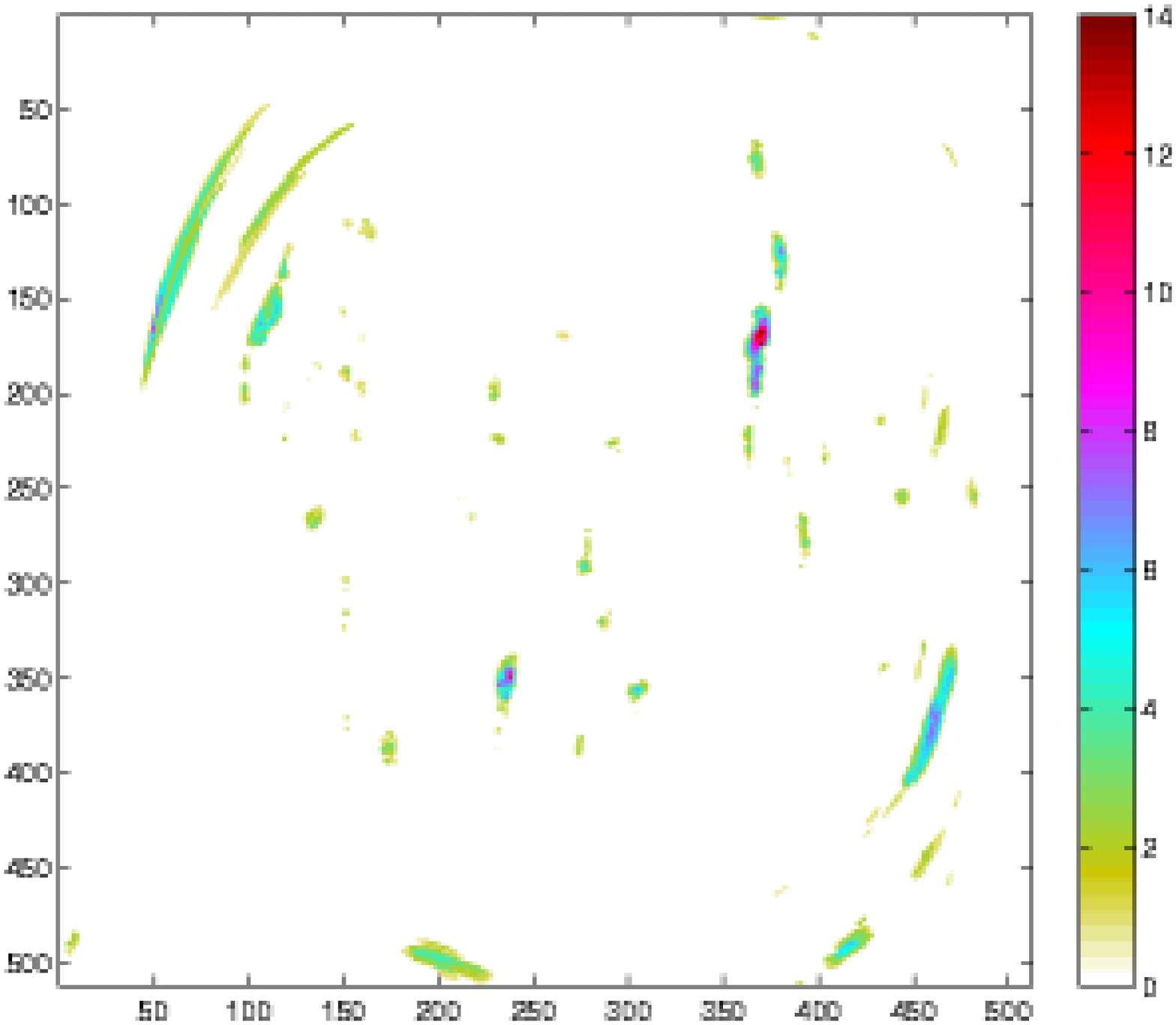}\\$32\times 64$
\end{minipage}
\begin{minipage}{5.60cm}
 \epsfxsize=5.5cm
 \epsfbox{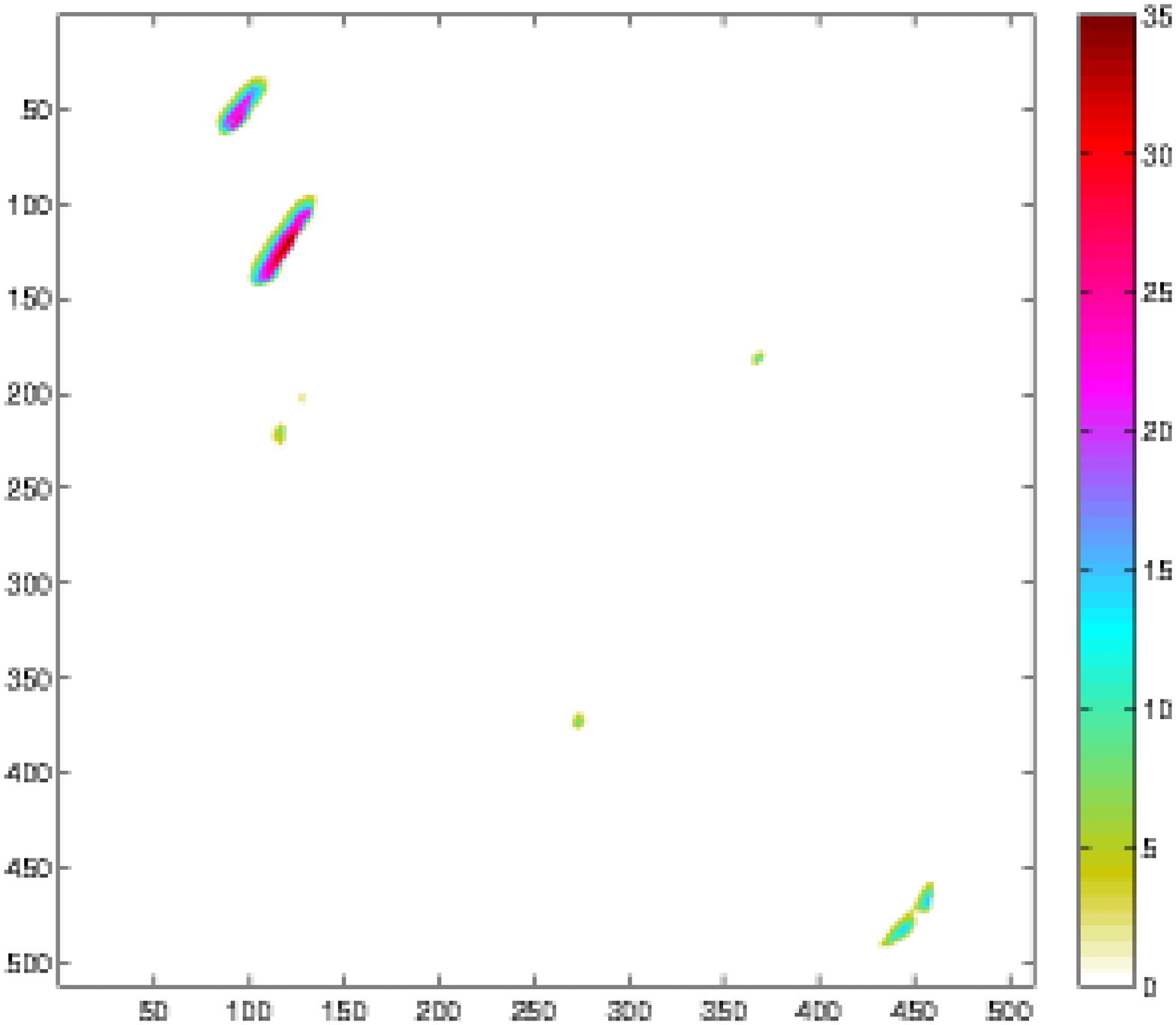}\\$32\times 128$
\end{minipage}
\caption{Finding the most similar part using the algorithm.
The test image has size of $54\times 145$. 
The size of the rectangle block $S$ is shown for each image. 
}\label{condimg}
\end{center}
\end{figure*}

We can further refine the results of the previous section by 
fixing the global threshold $T$ in a way to have only 
some fraction of the excited neurons. 
Thus we obtain a bump activity of the 
network, previously considered in \cite{EK3,EK4,AleYasser1}.
A sample result is shown in Fig.\ref{figbump}.

Regarding the time analysis of the procedure, one can see that 
the execution times are 
proportional to the number of the weights $w$. Having in mind that 
the connectivity is between the blocks, and that we can use a 
fraction of blocks less than $1/N^d$, the execution time can drop to order inferior
to the $N^{2d}$ limit. Thus, the number of steps to achieve the attractor is of 
order $d \log N$.{  It does not grow faster with the dimension than in a linear manner.}

\subsection{Conditional distribution}

A case of special practical interest is to find the most unusual part of the image with respect to some database of images or with
respect to a single image. 
Therefore, we are looking for the conditional probability of the occurrence of the blocks with respect to that database/image.
We can impose that conditional distribution by using the network constructed in the previous section. 
However, if we look for the dependence and the conditional distribution of only one test image $A_p$, we can do it in a easier way, namely we can take from the image $A_p$ the patterns, $\{X\}$ with a shape $S$ in a random manner.
That means to change step 1 of the algorithm to the following:

1c. Select at random a point of $A_p$ as an origin of the shape $S$. Use this area as a projection operator (normalizing and subtracting the mean brightness).

\begin{figure}[th]
\begin{center}
\begin{minipage}{3.0cm}
 \epsfxsize=2.0cm
 \epsfbox{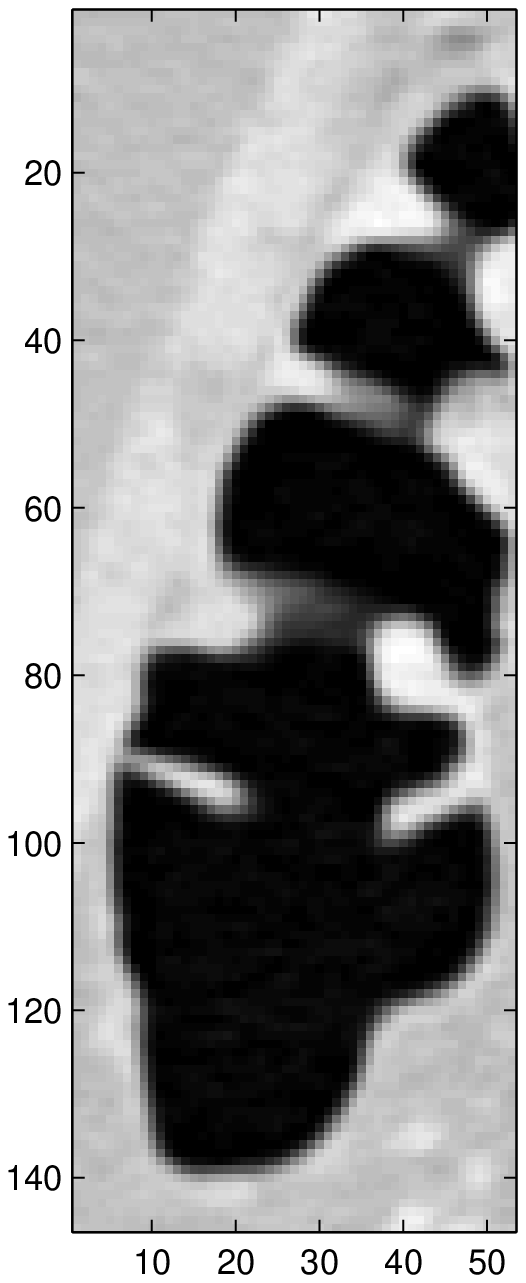}
\end{minipage}
\caption{The figure of patterns we use to find similar and dissimilar parts of the image.}\label{modelcolon}
\end{center}
\end{figure}

Then we can answer the following two questions at the same time - What is the part of the image $A$ with shape $S$ that is most similar to the image $A_p$ and what is the part of the image $A$ most dissimilar to the parts of the image $A_p$? The first answer is related to pattern recognition problem.  As an example, if we look for the colon in the CT image shown in Fig.\ref{section_48}, the test image can look like Fig.\ref{modelcolon}.

If the shape $S$ is small, then the statistics would be more or less universal and we cannot expect that the result would be very specific to the image $A_p$. If we increase the size of the shape $S$, the result will be more and more specific. 
If the size of the $S$ is similar to the size of $A_p$, we can expect highly specific response.
We have found empirically that in order to achieve satisfactory result we must use $\gamma \approx 1$, e.g. to eliminate the dependence on the contrast. 

The results of the conditioning are shown in Fig.\ref{condimg}. 
We condition one image of the colon in Fig. 3 to the one in Fig.12. We find that the recognition is very good. It does not depend on the dimensionality of the image.  
The first two panels of Fig.\ref{condimg} with smallest $S$, $16\times 16$ and $32\times 32$, 
have strong mixture between negative and positive large projections (not shown in the figure). It accentuates more or less the borders, but with mixed sign of the projections. The last image has only positive correlations with the test image in the upper left corner. The position of the colon (See Fig.\ref{section_48}) is detected correctly.

Because the methods of using Hebb-like network, described in the previous section is independent of the one, described in this section, we can combine them by simply applying them together.
Actually we need to connect only the points corresponding to the $M$ blocks of  $A_p$, that we have selected in step 1c, to compute the projections. Note that we used only 30 samplings in the previous experiments. 
{  The effect of the network application is pruning of 
the spurious part of the images, especially with big size of $S$.}

\section{Discussion and Future Directions}

In this paper we present a method to find the most unusual part in 
two and higher dimensional images, when its shape is fixed, but in general arbitrary. 
The method is almost independent on the size of the shape in terms of the execution speed and time. 
It gives good results on experimental images without 
predefined model of the interesting event.

{  The method works equally well for 2D and 3D images. 
It is also fast enough using 3D images.}

One necessary future development of the algorithm is to achieve practical 
and computable criteria of the "rareness" of the block and 
comparing the results on large enough database in order to have qualitative 
measure of the results.
The criterion must be different from Eq. (1),  
because its direct computing tends to be very slow and unstable. 

It is possible to speed up the process by using multi-resolution approach.
For example, we can use the downsized images and after that we can search in the original images only in the points detected in the smaller images.
However, this approach needs future exploration, because in its naïve version it can be used only when the convolution in space domain is faster than the convolution in frequency domain.


Among the future applications of the present method, one
could mention the achievement of
experiments on different type of images and large image databases and 
experiments on acceleration of the network due to the special equivalence
class construction. 

\section{Acknowledgments}
 The authors thank for hospitality and financial support
the Abdus Salam International Center for Theoretical Physics, Trieste, Italy,
where the part concerning the analysis of three-dimensional images
has been performed.
The work is also financially supported by Spanish Grants
TIN 2004--07676-G01-01, TIN 2007--66862 (K.K.) and 
DGI.M.CyT.FIS2005-1729 (E.K.).

\end{document}